\documentclass{moriond}

\usepackage{amssymb,amsmath}
\def\be{\begin{equation}}
\def\ee{\end{equation}}
\def\beqn{\begin{eqnarray}}
\def\eeqn{\end{eqnarray}}
\def\no{\nonumber}
\def\ba{\begin{array}{c}}
\def\bat{\begin{array}{ccc}}
\def\ea{\end{array}}
\def\bi{\begin{itemize}}
\def\ei{\end{itemize}}
\newcommand{\ta}[0]{\tilde \alpha}

\def\be{\begin{equation}}
\def\ee{\end{equation}}
\def\bea{\begin{eqnarray}}
\def\eea{\end{eqnarray}}

\begin{document}
\vspace*{4cm}
\title{On the search for a second scalar doublet at the LHC}

\author{ALEJANDRO CELIS}

\address{Ludwig-Maximilians-Universit\"at M\"unchen, 
   Fakult\"at f\"ur Physik,\\
   Arnold Sommerfeld Center for Theoretical Physics, 
   80333 M\"unchen, Germany \vspace{0.2cm}} 

\maketitle\abstracts{
Motivated by the principle of natural flavour conservation, searches for signatures of a second scalar doublet at the LHC are usually limited to models with a very restricted Yukawa structure.   Strong correlations between the scalar couplings to fermions are fixed in these models due to the underlying flavour principle, making the results obtained from such analyses highly model dependent.     Moreover, constraints derived from flavour experiments also vary radically within the different models considered.     The hypothesis of Yukawa alignment provides a suitable framework to perform a more general analysis of $125$~GeV boson data and searches for additional scalars at the LHC, allowing for sizable new physics effects to be observed in the scalar sector while accommodating the stringent limits from flavour experiments at the same time.   This general setting includes the different models with natural flavour conservation as particular cases.    }

\section{Introduction}

This talk is aimed to experimentalists following the spirit of the ``Rencontres de Moriond" conference.     The discovery of a $125$~GeV SM-like scalar boson at the LHC can be interpreted as a confirmation that a complex scalar doublet is responsible for the spontaneous breaking of the electroweak (EW) gauge symmetry.      Three degrees of freedom of this doublet give mass to $W^{\pm}$ and $Z$, the remaining degree of freedom should be the observed $125$~GeV boson.  Of course, none of the fundamental principles of the SM fix the particle content of the theory and we should search, among other things, for possible signals of additional scalar fields.       I am concerned here with a simple extension of the SM by a second complex scalar doublet.    Such a simple extension gives rise to a rich phenomenology at flavour factories and collider experiments like the LHC.    By extending the SM in such a way, we are not aiming to provide a more fundamental theory of Nature than the SM, the main motivation is to analyze what experimental data is telling us about the EW symmetry breaking sector.    On the other hand, an extended scalar sector as the one analyzed here could very well be regarded as a low energy effective theory of a more fundamental theory.

By extending the SM with a second scalar doublet $\Phi_i$ ($i=1,2$), we obtain a scalar spectrum composed of three neutral scalars ($\varphi^0_i =\{h, H, A\}$ ) and a charged scalar ($H^{\pm}$). One of the neutral scalars of the theory should correspond to the observed $125$~GeV state while the additional scalar states would have been missed so far in experimental searches.      Searches for signatures of a second scalar doublet at the LHC are strongly influenced by the observed strong suppression of flavour changing neutral currents (FCNCs).   A second scalar doublet introduces in general new sources of flavour violation into the theory giving rise to tree-level FCNCs in the scalar sector.   FCNCs are strongly constrained by low-energy flavour experiments, motivating the addition of an underlying mechanism which suppresses such dangerous terms.   A well known mechanism to achieve this is the principle of natural flavour conservation (NFC),\cite{Glashow:1976nt,Paschos:1976ay} which guarantees the absence of FCNCs at tree-level by assuming that the Lagrangian is invariant under a discrete $\mathcal{Z}_2$ symmetry.      The principle of NFC can be realized in different ways, giving rise to 4 different models, known as type I, II, X, Y.    Each of these models have different correlations among the scalar couplings to fermions.

The aim of my talk is to discuss the possibility to search for a second scalar doublet in a more general way.   Of particular relevance for LHC studies are the scalar couplings to the third generation fermions.  We would like to have a framework where the scalar couplings to $(t,b,\tau)$ are all independent a priori.  If there is actually any relation between these couplings it should be made manifest as a result of the experimental analyses.     Keeping the most general Yukawa structure of the model with two scalar doublets is not a good option since it contains too many free parameters, most of which are already strongly constrained by flavour bounds.      A convenient framework should incorporate the information obtained from flavour physics experiments while still allow for sizable effects to be observed at the LHC.    Based on this premises, it could be argued that a predictive, practical and general model for the search of an additional scalar doublet at the LHC should have the following features:

%%%
\begin{itemize}     
\item It should allow for independent deviations in the scalar couplings to the third generation fermions $(t, b, \tau)$.   
\item Flavour changing neutral currents in the scalar sector should be absent or strongly suppressed.         
\item It should be possible to implement constraints from flavour physics in a consistent manner without enlarging the number of free parameters in the analysis.  
\item Models with natural flavour conservation (type I, II, X, Y) should be recovered as particular cases.   
\end{itemize}
%%%%

The hypothesis of Yukawa alignment provides precisely the desired framework.\cite{Pich:2009sp}   Flavour constraints on the model still allow sizable effects associated to the scalar sector to be observed at the LHC.\cite{Jung:2010ik}$^{-}$\cite{Ilisie:2014hea}      An additional simplification comes from assuming that CP is conserved in the scalar sector and the only source of CP-violation is the phase of the CKM matrix as in the SM, this assumption can be justified by the non-observation of electric dipole moments.    A concise and self-contained description of the main aspects of this framework is given here.        In Sec.~\ref{sec:model} the extension of the scalar sector of the SM by a second scalar doublet is discussed.    Limits of the parameter space in which one of the neutral scalars has SM-like properties are presented in Sec.~\ref{SMH}.         Finally in Sec.~\ref{Yukasec} the hypothesis of Yukawa alignment is introduced and it is shown that the different models with NFC correspond to specific cases of this framework.

\section{Adding a second scalar doublet  \label{sec:model}}
The scalar doublets can be parametrized in full generality by
\begin{equation}  \label{Higgsba}
\Phi_1=\left[ \begin{array}{c} G^+ \\ \frac{1}{\sqrt{2}}\, (v+S_1+iG^0) \end{array} \right] \; ,
\qquad\qquad\qquad
\Phi_2 = \left[ \begin{array}{c} H^+ \\ \frac{1}{\sqrt{2}}\, (S_2+iS_3)   \end{array}\right] \; .
\end{equation}
Here we have chosen a scalar basis in which only one of the scalar doublets acquires a non-vanishing vev $v = (\sqrt{2} G_F )^{-1/2} \simeq 246$~GeV.   The most general CP-conserving scalar potential compatible with the SM gauge symmetry is 
\beqn\label{eq:pote} 
V & = & \mu_1\; \Phi_1^\dagger\Phi_1\, +\, \mu_2\; \Phi_2^\dagger\Phi_2 \, +\, \left[\mu_3\; \Phi_1^\dagger\Phi_2 \, +  \mathrm{h.c.} \right]
\no\\ & + & \lambda_1\, \left(\Phi_1^\dagger\Phi_1\right)^2 \, +\, \lambda_2\, \left(\Phi_2^\dagger\Phi_2\right)^2 \, +\,
\lambda_3\, \left(\Phi_1^\dagger\Phi_1\right) \left(\Phi_2^\dagger\Phi_2\right) \, +\, \lambda_4\, \left(\Phi_1^\dagger\Phi_2\right) \left(\Phi_2^\dagger\Phi_1\right)
\no\\ & + & \left[  \left(\lambda_5\; \Phi_1^\dagger\Phi_2 \, +\,\lambda_6\; \Phi_1^\dagger\Phi_1 \, +\,\lambda_7\; \Phi_2^\dagger\Phi_2\right) \left(\Phi_1^\dagger\Phi_2\right)
\, +\, \mathrm{h.c.}\right]\, ,
\eeqn
with real coefficients $\{\mu_j, \lambda_j\}$.     The quartic couplings are expected to be $\lambda_j \sim \mathcal{O}(1)$ and are assumed to be bounded by perturbativity.  The following relations are derived from the fact that the minimum of the potential is an extremum point: $\mu_1\; =\; -\lambda_1\, v^2$ and $\mu_3\; =\; - \lambda_6\, v^2/2$.  Introducing a second scalar doublet to the scalar sector of the SM implies the existence of a mass scale in the potential, $\mu_2$, not related a priori to the Fermi constant.      The masses of the physical scalars are given in terms of the two available mass scales $\mu_2$ and $v$, accompanied by quartic couplings.  The mass of the charged scalar is
\begin{align} \label{eq:mplus}
M_{H^\pm}^2\; =\; \mu_2 + \frac{1}{2}\,\lambda_3\, v^2 \,.
\end{align}
The neutral scalars are given by
\be \label{eq:CP_mixing}
\left(\ba h\\ H  \\ A \ea\right)\; = \;
\left[\bat \cos{\tilde\alpha} & \sin{\tilde\alpha} & 0  \\   -\sin{\tilde\alpha} & \cos{\tilde\alpha}  & 0 \\ 0 & 0 & 1 \ea\right]\;
\left(\ba S_1\\ S_2  \\ S_3\ea\right) \,,
\ee
with $M_{H} \geq M_h$ by convention.     The mixing angle is given by
\beqn \label{eq:wclimit}
   \sin  \tilde \alpha     &=&     \left( \frac{  2 \lambda_1 v^2 - M_h^2 }{     M_H^2 - M_h^2     } \right)^{1/2}   \,,   \qquad  \sin  \tilde \alpha   \cos \ta =    \frac{-  \lambda_6 v^2  }{  M_H^2 - M_h^2   }     \,.\eeqn
 We can restrict to $0  \leq \tilde \alpha \leq \pi$ since we are free to perform a phase redefinition of the CP-even fields. Furthermore, a global rephasing of the second scalar doublet $\Phi_2 \rightarrow - \Phi_2$ is unphysical.\cite{Davidson:2005cw}   We can characterize this two-fold ambiguity in terms of the sign of $\lambda_6$, without loss of generality we can then fix the sign of $\lambda_6$.  By convention we choose $\lambda_6 \leq 0$ which implies $0  \leq \tilde \alpha \leq \pi/2$.   The masses of the neutral scalars are 
\be \label{eq:CPC_mass}
M_h^2\; =\;\frac{1}{2}\,\left( \Sigma-\Delta\right)\, ,
\qquad
M_H^2\; =\;\frac{1}{2}\,\left( \Sigma+\Delta\right)\, ,
\qquad
M_A^2 \; =\; M_{H^\pm}^2\, +\, v^2\,\left(\frac{\lambda_4}{2} - \lambda_5\right)\, ,
\ee
with
\beqn\label{eq:Sigma}
\Sigma & =&      M_{H^{\pm}}^{2}   +  \left( 2 \lambda_1 +  \frac{ \lambda_4 }{2}  + \lambda_5 \right)  v^2   \, , \qquad 
\Delta  =\sqrt{\left[  M_A^2 + 2  (\lambda_5 - \lambda_1) v^2    \right]^2 + 4 v^4 \lambda_6^2}\, .
\eeqn
The scalar potential of the two-doublet model is parametrized in terms of 8 real parameters $\{\mu_2, \lambda_k\}$ ($k=1,\ldots, 7$).   It is possible to trade some of these parameters for other quantities which are more closely related to physical observables as $\{M_h, M_H, M_A, M_{H^{\pm}}, \tilde \alpha, \lambda_2, \lambda_3, \lambda_7 \}$:
\begin{align}  \label{recon}
\lambda_1 &=   \frac{1}{2 v^2}  \left[    M_h^2 \, \cos^2 \tilde \alpha + M_H^2 \, \sin^2 \tilde \alpha           \right] \,,  \qquad \lambda_4 =   \frac{1}{v^2} \left[  M_h^2 \, \sin^2 \tilde \alpha + M_H^2 \, \cos^2 \tilde \alpha + M_A^2 - 2 M_{H^{\pm}}^2     \right]   \,, \nonumber \\
\lambda_5 &=   \frac{1}{2 v^2} \left[  M_h^2 \, \sin^2 \tilde \alpha + M_H^2 \, \cos^2 \tilde \alpha - M_A^2     \right]  \,,  \qquad \lambda_6 = - \frac{1}{v^2} (  M_H^2 - M_h^2  )  \cos \tilde \alpha \, \sin \tilde \alpha \,.
\end{align}
The number of free parameters in the scalar potential gets effectively reduced to 7 due to the measurement of the scalar boson mass at the LHC.  Since the CP-odd state $A$ does not couple to the massive gauge vector bosons, we know that $M_{h}$ or $M_H$ should correspond to $125$~GeV.   The neutral scalar couplings to the massive gauge vector bosons are again determined by the mass generation mechanism up to mixing effects
\beqn\label{eq:Lphv2}
\mathcal{L}_{\varphi V^2} & = &  \frac{2}{v}\;   \left( \cos{\tilde \alpha}  \,    h     -\sin{\tilde \alpha} \, H     \right)  \,\left[ M_W^2\, W^\dagger_\mu W^\mu  +  \frac{1}{2}\, M_Z^2\, Z_\mu Z^\mu \right]\, .
\eeqn
Note that the relation $M_W/(M_Z \cos \theta_W) =1$ remains valid at tree-level.   The interactions of the scalar fields with fermions are not directly correlated anymore to the fermion masses
\beqn\label{Ilagrangianl}
 \mathcal L_{Y} &= &  - \, \sum_{\varphi^0_i, f= u, d, \ell} \,   \varphi^0_i \, \bar{f}\,  Y_{f}^{\varphi^0_i} \,  \mathcal{P}_R \, f  \nonumber \\
\; && - \frac{\sqrt{2}}{v}    \, H^{+}   \, \left\{    \bar u \left[   V_{\mbox{\scriptsize{CKM}}} \, \Pi_{d} \,\mathcal{P}_R - \Pi_{u}^{\dag}\,  V_{\mbox{\scriptsize{CKM}}} \, \mathcal{P}_L     \right] d \,   + \bar \nu \, \Pi_{\ell} \,\mathcal{P}_R \, \ell \,    \right\}  +    \;\mathrm{h.c.}   \,,
\eeqn
with
\begin{align}    \label{Ilaggen}
v \,Y_{d, \ell}^{\varphi^0_i} &=  M_{d, \ell} \, \mathcal{R}_{i1} + \Pi_{d, \ell} \, \left( \mathcal{R}_{i2} + i \,\mathcal{R}_{i3}  \right)   \,,  \qquad 
v \,Y_{u}^{\varphi^0_i} =    M_{u} \, \mathcal{R}_{i1} + \Pi_{u} \, \left( \mathcal{R}_{i2} -  i \,\mathcal{R}_{i3}  \right) \,.
\end{align}
Here $\varphi^0_i= \mathcal{R}_{ij} S_j$ with the mixing matrix $\mathcal{R}$ defined in Eq.~\ref{eq:CP_mixing}.   The $M_{f=u,d,l}$ are the diagonal fermion mass matrices while the $\Pi_{f=u,d,l}$ represent arbitrary real matrices in flavour space, giving rise to tree-level FCNCs in the scalar sector.  The chiral projectors $\mathcal{P}_{L,R} = (1 \mp \gamma_5)/2$ are denoted as usual and  $V_{\mbox{\scriptsize{CKM}}}$ is the CKM matrix.

\section{A SM-like scalar boson at $\mathbf{125}$~GeV \label{SMH} }
The observation of a SM-like boson at $125$~GeV has important implications for the structure of the scalar sector presented previously.    The properties of the scalar sector depend strongly on the hierarchy between $\mu_2$ and the EW scale.  For $\mu_2 \gg v^2$ and perturbative quartic couplings the second scalar doublet $\Phi_2$ decouples from the theory, leaving a SM-like scalar sector at the EW scale.\cite{Gunion:2002zf}      The scalar masses are given in this limit by 
\be
 M_h^2 \simeq 2 \lambda_1 v^2 + \mathcal{O}\left(\dfrac{v^4}{\mu_{2}}\right) \,, \qquad\qquad
  M_{H}^2 \simeq M_A^2 \simeq  M_{H^{\pm}}^2 = \mu_2 + \mathcal{O}(v^2)  \,.
\ee
The mixing between the two scalar doublets is suppressed by the high mass scale, $\tan \ta  \simeq \mathcal{O}( v^2/\mu_{2}) $, so that the couplings of the light CP-even scalar boson with vector bosons and fermions approach the SM values
\be
\cos \ta = ( 1 + \tan^2 \ta )^{-1/2} \simeq 1 +  \mathcal{O}\left(   \dfrac{v^4}{ \mu_{2}^2}  \right)  \,, \qquad\qquad
 Y_f^{h}  \simeq \frac{M_{f}}{v}  + \mathcal{O}\left(    \dfrac{ v^2 }{  \mu_{2}  }  \right)   \,.
\ee
It is said that the approach to the decoupling is faster for the scalar coupling to massive vector bosons than to fermions.\cite{Gunion:2002zf}  In such decoupling scenario, obviously only the lightest CP-even state $h$ could play the role of the $125$~GeV boson.     In models where the second mass scale $\mu_2$ is absent or is related to the EW scale $\mu_2 \sim v^2$, due to an underlying symmetry for example, there is no decoupling.\cite{Gunion:2002zf}   Even in this scenario it is possible to have a light SM-like scalar boson.  This occurs when $\lambda_6 \rightarrow 0$ since the mixing between the two scalar doublets vanishes in this limit.\cite{Gunion:2002zf}     In this case the additional scalars should also be around the EW scale assuming perturbativity of the quartic scalar couplings.     In this limit any of the CP-even states could play the role of the SM-like $125$~GeV boson:

\begin{itemize}

\item  For $\lambda_6 \rightarrow 0$ and $M_A^2 + 2 (\lambda_5 - \lambda_1) v^2 > 0$ we get $\cos 2  \tilde \alpha \simeq 1 + \mathcal{O}(\lambda_6^2)$, so that $\tilde \alpha = 0 + 1/2 | \arccos \left( \mathcal{O}(\lambda_6^2)  \right) |$.  The light CP-even state $h$ becomes SM-like $(\cos \ta \simeq 1)$.

\item  For $\lambda_6 \rightarrow 0$ and $M_A^2 + 2 (\lambda_5 - \lambda_1) v^2 < 0$ we get $\cos 2 \tilde \alpha \simeq - 1 + \mathcal{O}(\lambda_6^2)$, so that $\tilde \alpha = \pi/2 - 1/2 |\arccos \left( \mathcal{O}(\lambda_6^2)  \right) |$.  The heavy CP-even state $H$ becomes SM-like $(\sin \ta \simeq 1)$.
\end{itemize}

Here we have used the following exact identity
\be
\cos 2 \ta =  \frac{ M_A^2 + 2 (\lambda_5 - \lambda_1) v^2     }{  \sqrt{  \left[  M_A^2 + 2  (\lambda_5 - \lambda_1) v^2    \right]^2 + 4 v^4 \lambda_6^2  } } \,.
\ee

\section{A framework for the search of a second scalar doublet at the LHC \label{Yukasec}}
From the phenomenological point of view we are interested in a scenario with additional scalar fields accessible at the LHC and with possible deviations from the SM in the $125$~GeV boson couplings.  The Yukawa alignment hypothesis provides a suitable framework to interpret LHC data in such scenario.   The Yukawa alignment condition is based on the assumption that the Yukawa matrices for each type of fermion are aligned in flavour space, see Eq.~\ref{Ilagrangianl},\footnote{Though the Yukawa alignment condition is not stable under quantum corrections, accidental flavour symmetries in the Lagrangian protect the flavour structure of the model.   For the phenomenological purposes we are interested the Yukawa aligned structure can be considered stable.\cite{Pich:2009sp,Jung:2010ik,Braeuninger:2010td,Li:2014fea,Abbas:2015cua}  }
\be
\Pi_{d,l} = \varsigma_{d,l} M_{d,l} \,, \qquad  \Pi_u= \varsigma_u M_u  \,.
\ee
Here the flavour universal alignment parameters $\varsigma_{f=u,d,l}$ are arbitrary real numbers (assuming CP-conservation).  The Yukawa Lagrangian reads
\beqn\label{lagrangian}
 \mathcal L_Y & = &  - \frac{\sqrt{2}}{v}\; H^+ \left\{ \bar{u} \left[ \varsigma_d\, V_{\mbox{\scriptsize{CKM}}} M_d \mathcal P_R - \varsigma_u\, M_u V_{\mbox{\scriptsize{CKM}}} \mathcal P_L \right]  d\, + \, \varsigma_l\, \bar{\nu} M_l \mathcal P_R l \right\}
\nonumber \\
& & -\,\frac{1}{v}\; \sum_{\varphi^0_i, f=u,d,l}\, y^{\varphi^0_i}_f\, \varphi^0_i  \; \left[\bar{f}\,  M_f \mathcal P_R  f\right]
\;  + \;\mathrm{h.c.} 
\eeqn
The fermionic couplings of the neutral scalar fields are given, in units of the SM scalar couplings by
\begin{align}  \label{equations1intro}
&& y_{f}^h & = \cos{\tilde\alpha} + \varsigma_f \sin{\tilde \alpha} \!\ , &&& y_{d,l}^A & =  i\,\varsigma_{d,l}  \!\ , &&  \notag \\
& & y_{f}^H & = -\sin{\tilde\alpha} + \varsigma_f \cos{\tilde \alpha} \!\ , &&&
y_{u}^A \; & =\; -i\, \varsigma_u  \!\  \,.
\end{align}
The model contains 10 free parameters, 7 coming from the scalar potential as explained in Sec.~\ref{sec:model} and 3 from the Yukawa sector  $\{ \varsigma_u, \varsigma_d, \varsigma_lÊÊ\}$.      This framework satisfies all the conditions spelled in the introduction for the search of a second scalar doublet at the LHC.   The first three conditions are manifest in Eqs.~\ref{lagrangian} and \ref{equations1intro}.     It can also be shown that models with NFC are recovered as particular limits of the Yukawa aligned model. 

\subsection{Models with natural flavour conservation}
NFC models are usually expressed in a scalar basis $\phi_i$ $(i=1,2)$ where both doublets acquire vevs $\langle \phi_i^0 \rangle = v_i/\sqrt{2}$, with $v^2 =  v_1^2 + v_2^2$.  Such basis is related to the one in Eq.~\ref{Higgsba} by an orthogonal transformation parametrized by $\tan \beta \equiv v_2/v_1$.  
The usual convention for models with NFC is to use the rephasing freedom $\Phi_2 \rightarrow - \Phi_2$ to fix the sign of $\tan \beta \geq 0$, or equivalently $0  \leq   \beta \leq \pi/2$.     By doing so, we can no longer fix the sign of $\lambda_6$ so that $0 \leq \tilde \alpha  \leq \pi$.

%%%%%%%%%%%%%%%%%%%%%%%%%%%%%%%%%%%%%%%%%%%%%%s %%%%%%%%%%%%%%%%%%%%%%%%%%%%%%%%%%%%%%%%%%%%%%%%%%%
\begin{table}[h]\begin{center}
\caption{\it \small Models with natural flavour conservation.}
\vspace{0.1cm}
\begin{tabular}{|c|c|c|c|}
\hline
Model & $\varsigma_d$ & $\varsigma_u$ & $\varsigma_l$  \\
\hline
Type I  & $\cot{\beta}$ &$\cot{\beta}$ & $\cot{\beta}$ \\
Type II & $-\tan{\beta}$ & $\cot{\beta}$ & $-\tan{\beta}$ \\
Type X (lepton-specific) & $\cot{\beta}$ & $\cot{\beta}$ & $-\tan{\beta}$ \\
Type Y  (flipped) & $-\tan{\beta}$ & $\cot{\beta}$ & $\cot{\beta}$ \\
\hline
\end{tabular}
\label{tab:models}
\end{center}\end{table}
%%%%%%%%%%%%%%%%%%%%%%%%%%%%%%%%%%%%%%%%%%%%%%%%%%%%%%%%%%%%%%%%%%%%%%%%%%%%%%%%%%%%%%%%%%%%%%%%%%%%%%%%%%%%%%%%%%%%%%%

NFC can be obtained by considering a softly-broken $\mathcal{Z}_2$ symmetry under which $\phi_1$ is even and $\phi_2$ is odd.   Four different NFC models are obtained depending on the assigned fermionic charges (types I, II, X, Y).    The presence of such a symmetry in the theory imply correlations among the parameters of the model.     In models with NFC all the alignment parameters are given in terms of $\tan \beta$, see Table~\ref{tab:models}.  These relations imply strong correlations between the scalar boson couplings to fermions.   For example, taking the type I model we get $|y_{u}^{\varphi^0_i}| = |y_{d}^{\varphi^0_i}|  = |y_{l}^{\varphi^0_i}|$ from Eq.~\ref{equations1intro}.   

Terms with an odd parity under the $\mathcal{Z}_2$ symmetry are forbidden in the scalar potential, translating into correlations among the parameters of the scalar potential
\begin{align} \label{ntc}
 \lambda^{\prime}_6 & =     ( \lambda_1 - \lambda_3 - \lambda_5  ) \, c_{\beta}^3 s_{\beta} + ( \lambda_4 + \lambda_5 - \lambda_2    )\,  c_{\beta} s_{\beta}^3   + \lambda_6 \, ( c_{\beta}^4 -  c_{\beta}^2 s_{\beta}^2) + 2 \lambda_7\, c_{\beta}^2 s_{\beta}^2 = 0   \,, \nonumber \\
\lambda_7^{\prime} &=   ( \lambda_1 - \lambda_3 - \lambda_4 - \lambda_5) \,c_{\beta} s_{\beta}^3 + ( \lambda_5 -\lambda_2 ) \,c_{\beta}^3 s_{\beta}   + 2 \lambda_6 \, s_{\beta}^2 c_{\beta}^2 + \lambda_7\, ( c_{\beta}^4 - s_{\beta}^2 c_{\beta}^2 )  = 0  \,. 
\end{align}
Here $s_{\beta} \equiv \sin \beta$ and $c_{\beta} \equiv  \cos \beta$.  The parameters $\lambda^{\prime}_{6,7}$ are the coefficients of $(\phi_1^{\dag}  \phi_1 ) (\phi_1^{\dag}  \phi_2)$ and $(\phi_2^{\dag}  \phi_2 ) (\phi_1^{\dag}  \phi_2)$ in the scalar potential.    If the $\mathcal{Z}_2$ symmetry is an exact symmetry of the Lagrangian (it is not softly broken) an additional relation appears between the scalar potential parameters:
\be \label{ntc2}
m_{12}^{\prime \, 2} = ( \mu_2  - \mu_1 ) c_{\beta} s_{\beta} - \mu_3 c_{2 \beta} = 0 \,,
\ee
with $m_{12}^{\prime \, 2 }$ being the coefficient of the term $\phi_{1}^{\dag} \phi_2$ in the scalar potential.    Note that there is no decoupling limit in this scenario since the exact $\mathcal{Z}_2$ symmetry relates $\mu_2$ to the EW scale.\cite{Gunion:2002zf}

\section{Conclusions}

Signatures of a second scalar doublet could be within the reach of the next runs of LHC.   Current analyses of $125$~GeV boson data and direct searches for additional scalars at the LHC are being performed by the experimental collaborations within models with natural flavour conservation, which have a very restricted Yukawa structure.   Ideally, we should search for signatures of a second scalar doublet in a more general way.   The Yukawa aligned model provides a predictive, practical and general framework for this purpose.      Allowing for sizable new physics effects associated to the scalar sector to be observed at the LHC while satisfying the stringent flavour limits.     Models with natural flavour conservation are recovered as particular cases.

\section*{Acknowledgments}
I thank the organizers of ``The 50th Rencontres de Moriond, Electroweak Session".   I am grateful to V.~Ilisie and A.~Pich for useful comments on this manuscript.   A.C. is supported by the Alexander von Humboldt Foundation.


\section*{References}

\begin{thebibliography}{99}


 


\bibitem{Glashow:1976nt}
  S.~L.~Glashow and S.~Weinberg, Phys.\ Rev.\ D {\bf 15} 1958 (1977).
  
\bibitem{Paschos:1976ay}
E.~A.~Paschos, Phys.\ Rev.\ D {\bf 15} 1966 (1977).
  




  
  
\bibitem{Pich:2009sp}
  A.~Pich and P.~Tuzon, Phys.\ Rev.\ D {\bf 80} 091702 (2009). 
  
  


\bibitem{Jung:2010ik}
  M.~Jung, A.~Pich and P.~Tuzon, JHEP {\bf 1011} 003 (2010).
 
 %\cite{Cervero:2012cx}
\bibitem{Cervero:2012cx}
  E.~Cervero and J.~M.~Gerard,
  %``Minimal violation of flavour and custodial symmetries in a vectophobic Two-Higgs-Doublet-Model,''
  Phys.\ Lett.\ B {\bf 712} 255 (2012).
 % [arXiv:1202.1973 [hep-ph]].
  %%CITATION = ARXIV:1202.1973;%%
  %39 citations counted in INSPIRE as of 24 Apr 2015
  
\bibitem{Altmannshofer:2012ar}
  W.~Altmannshofer, S.~Gori and G.~D.~Kribs,
  %``A Minimal Flavor Violating 2HDM at the LHC,''
  Phys.\ Rev.\ D {\bf 86} 115009 (2012).
  
\bibitem{Bai:2012ex}
  Y.~Bai {\it et al.}
  %``General two Higgs doublet model (2HDM-G) and Large Hadron Collider data,''
  Phys.\ Rev.\ D {\bf 87} 115013 (2013).
  
  \bibitem{Celis:2013rcs}
  A.~Celis, V.~Ilisie and A.~Pich, JHEP {\bf 1307} 053 (2013).
  

  
  
\bibitem{Lopez-Val:2013yba}
  D.~Lopez-Val, T.~Plehn and M.~Rauch, JHEP {\bf 1310} 134 (2013).
  


  

\bibitem{Barger:2013ofa}
  V.~Barger {\it et al.}
  %``Scrutinizing the 125 GeV Higgs boson in two Higgs doublet models at the LHC, ILC, and Muon Collider,''
  Phys.\ Rev.\ D {\bf 88} 11,  115003 (2013).
  
  
  
  
\bibitem{Celis:2013ixa}
  A.~Celis, V.~Ilisie and A.~Pich, JHEP {\bf 1312}  095 (2013).
 
 %\cite{Wang:2013sha}
\bibitem{Wang:2013sha}
  L.~Wang and X.~F.~Han,
  %``Status of the aligned two-Higgs-doublet model confronted with the Higgs data,''
  JHEP {\bf 1404} 128 (2014).
  
  %\cite{Ilisie:2014hea}
\bibitem{Ilisie:2014hea}
  V.~Ilisie and A.~Pich,
  %``Low-mass fermiophobic charged Higgs phenomenology in two-Higgs-doublet models,''
  JHEP {\bf 1409}  089 (2014).


  
 
  
  %\cite{Davidson:2005cw}
\bibitem{Davidson:2005cw}
  S.~Davidson and H.~E.~Haber, Phys.\ Rev.\ D {\bf 72} 035004 (2005).
  %``Basis-independent methods for the two-Higgs-doublet model,''
   % [hep-ph/0504050].
  %%CITATION = HEP-PH/0504050;%%
  %145 citations counted in INSPIRE as of 24 Apr 2015
  



\bibitem{Gunion:2002zf}
  J.~F.~Gunion and H.~E.~Haber,
  %``The CP conserving two Higgs doublet model: The Approach to the decoupling limit,''
  Phys.\ Rev.\ D {\bf 67} 075019 (2003).
  
  
  %\cite{Braeuninger:2010td}
\bibitem{Braeuninger:2010td}
  C.~B.~Braeuninger, A.~Ibarra and C.~Simonetto,
  %``Radiatively induced flavour violation in the general two-Higgs doublet model with Yukawa alignment,''
  Phys.\ Lett.\ B {\bf 692} 189 (2010).


  %\cite{Li:2014fea}
\bibitem{Li:2014fea}
  X.~Q.~Li, J.~Lu and A.~Pich,
  %``$B_{s,d}^0 \to \ell^+\ell^-$ Decays in the Aligned Two-Higgs-Doublet Model,''
  JHEP {\bf 1406} 022 (2014).
  %%CITATION = ARXIV:1404.5865;%%
  %15 citations counted in INSPIRE as of 24 Apr 2015
  
  
  
  %\cite{Abbas:2015cua}
\bibitem{Abbas:2015cua}
  G.~Abbas, A.~Celis, X.~Q.~Li, J.~Lu and A.~Pich,
  %``Flavour-changing top decays in the aligned two-Higgs-doublet model,''
  arXiv:1503.06423 (2015).
  %%CITATION = ARXIV:1503.06423;%%
  
  



  

\end{thebibliography}
\end{document}